\documentclass[onecolumn]{article}
\usepackage{amsfonts}
\usepackage{amssymb}
\usepackage{amsmath}
\usepackage{dcolumn}

\input{tcilatex}

\begin{document}

\title{The Spacetime Algebra Approach to Massive Classical Electrodynamics
with Magnetic Monopoles}
\author{C. Cafaro\thanks{%
E-mail: carlocafaro2000@yahoo.it}, \ S. A. Ali\thanks{%
E-mail: alis@alum.rpi.edu} \\
Department of Physics, University at Albany-SUNY, \\
1400 Washington Avenue, Albany, NY 12222, USA}
\date{}
\maketitle

\begin{abstract}
Maxwell's equations with massive photons and magnetic monopoles are
formulated using spacetime algebra. It is demonstrated that a single
non-homogeneous multi-vectorial equation describes the theory. Two limiting
cases are considered and their symmetries highlighted: massless photons with
magnetic monopoles and finite photon mass in the absence of monopoles.
Finally, it is shown that the EM-duality invariance is a symmetry of the
Hamiltonian density (for Minkowskian spacetime) and Lagrangian density (for
Euclidean 4-space) that reflects the signature of the respective metric
manifold.
\end{abstract}

\section{Introduction}

Applications of Geometric Algebra (GA) to Maxwell's theory of
electromagnetism are known $\left[ 1\right] ,\left[ 2\right] ,\left[ 3\right]
,\left[ 4\right] $. In this paper the formalism of spacetime algebra is used
as in $\left[ 5\right] $ and $\left[ 6\right] $ to formulate classical
electrodynamics with massive photons and magnetic monopoles.

The layout of the paper is as follows. In section 2 a brief introduction to
the GA of spacetime is presented. In section 3.1 we describe the classical
field theory of our model, that is, electric charges interacting with Dirac
monopoles via massive photons or Proca fields, in the context of Lagrangian
dynamics. In section 3.2\textbf{\ }a formulation of Maxwell's theory of
electromagnetism with finite photon mass and magnetic monopoles is presented
and we demonstrate how to obtain a single nonhomogeneous multi-vectorial
equation describing the theory. In section 4\textbf{\ }we consider two
limiting cases of the general theory: i) massive electrodynamics in the
absence of magnetic monopoles and ii) standard electrodynamics with
monopoles. In the former case, the loss of local gauge invariance of the
theory is discussed. In the latter, the symmetry of the theory under duality
rotations is presented. In 5 we consider the invariances of the Lagrangian
and Hamiltonian densities, specifically under duality rotations and local
gauge transformations.

\section{An Outline of Spacetime Algebra}

The basic idea in geometric algebra is that of uniting the inner and outer
products into a single product, namely the \textit{geometric product}. This
product is associative and has the crucial feature of being invertible. The
geometric product between two spacetime vectors $a$ and $b$ is defined by%
\begin{equation}
ab=a\cdot b+a\wedge b\text{,}
\end{equation}%
where $a\cdot b$ is a scalar (a 0-grade multi-vector), while $a\wedge
b=i(a\times b)$ is a bi-vector (a grade-2 multi-vector). The quantity $i$ is
the unit \textit{pseudoscalar }which will be defined in (9), it is not the
unit imaginary number usually employed in physics. For convenience we recall
in brief the basic tools of GA that are utilized in our formulation of
massive classical electrodynamics in the presence of magnetic monopoles.

The GA of Minkowski spacetime is called \textit{spacetime algebra}. It is
generated by four orthogonal basis vectors $\left\{ \gamma _{\mu }\right\}
_{\mu =0..3}$ satisfying the relations%
\begin{equation}
\gamma _{\mu }\cdot \gamma _{\nu }=\frac{1}{2}\left( \gamma _{\mu }\gamma
_{\nu }+\gamma _{\nu }\gamma _{\mu }\right) \equiv \eta _{\mu \nu
}=diag(+---)\text{; }\mu \text{, }\nu =0..3
\end{equation}%
\begin{equation}
\gamma _{\mu }\wedge \gamma _{\nu }=\frac{1}{2}\left( \gamma _{\mu }\gamma
_{\nu }-\gamma _{\nu }\gamma _{\mu }\right) \equiv \gamma _{\mu \nu }\text{.}
\end{equation}%
We observe that (2) and (3) display the same algebraic relations as Dirac's $%
\gamma $-matrices. Indeed, the Dirac matrices constitute a representation of
the spacetime algebra. From (2) it is obvious that%
\begin{equation}
\gamma _{0}^{2}=1\text{, }\gamma _{0}\cdot \gamma _{i}=0\text{ and }\gamma
_{i}\cdot \gamma _{j}=-\delta _{ij}\text{; }i\text{, }j=1..3.
\end{equation}%
A basis for this 16-dimensional spacetime Clifford algebra $\mathfrak{cl}%
(1,3)$ is given by%
\begin{equation}
B=\left\{ 1,\gamma _{\mu },\gamma _{\mu }\wedge \gamma _{\nu },i\gamma _{\mu
},i\right\} \text{,}
\end{equation}%
whose elements represent scalars, vectors, bi-vectors, tri-vectors and
pseudoscalars respectively. A general multi-vector $M$ of the spacetime
algebra can be written as%
\begin{equation}
M=\dsum\limits_{k=0}^{4}\left\langle M\right\rangle _{k}=\alpha
+a+B+ib+i\beta \text{,}
\end{equation}%
where $\alpha $ and $\beta $ are real scalars, $a$ and $b$\ are real
spacetime vectors and $B$ is a bi-vector. A general spacetime vector $a$ can
be written as%
\begin{equation}
a=a^{\mu }\gamma _{\mu }\text{.}
\end{equation}%
By choosing $\gamma _{0}$ as the future-pointing timelike unit vector, the $%
\gamma _{0}$-vector determines a map between spacetime vectors $a$ and the
even subalgebra of the spacetime $\mathfrak{cl}^{+}\left( 1,3\right) $
algebra via the relation%
\begin{equation}
a\gamma _{0}=a_{0}+\vec{a}\text{,}
\end{equation}%
where $a_{0}=a\cdot \gamma _{0}$ and $\vec{a}\mathbf{=}a\wedge \gamma _{0}$.
Notice that the ordinary three-dimensional vector $\vec{a}$ can be
interpreted as a spacetime bi-vector $a\wedge \gamma _{0}$. The geometric
interpretation of these relations is the following: since a vector appears
to an observer as a line segment existing for a fixed period of time, it is
natural that what an observer perceives as a vector should be represented by
a spacetime bi-vector. Spacetime is a space of four dimensions with a
Lorentz signature, that is, $tr\left( \eta _{\mu \nu }\right) =-2$. The
metric $\eta _{\mu \nu }$ has one positive, three negative and no zero
eigenvalues. It is this property of the metric that distinguishes the
standard $\left( 3+1\right) $-dimensional space-time from the $4$%
-dimensional space of $SO(4)$, or the $\left( 2+2\right) $- dimensional
spacetime of $SO(2,2)$. The real Clifford algebra $\mathfrak{cl}(1,3)$ is
characterized by its total vector dimension $n=p+q$ $(n=4)$ and signature $%
s=p-q$ $\left( s=-2\right) $ where $p$ is the number of basis vectors with
positive norm $\left( p=1\right) $ and $q$ enumerates the basis vectors with
negative norm $\left( q=3\right) $. Many authors, especially general
relativists, use a Minkowski spacetime metric $+2$. This involves the
algebra $\mathfrak{cl}(3,1)$ where the spacelike vectors have positive norm.
The algebras $\mathfrak{cl}(1,3)$ and $\mathfrak{cl}(3,1)$ are not
isomorphic. In quantum field theory, for instance, a compelling reason to
choose $\mathfrak{cl}(1,3)$ over $\mathfrak{cl}(3,1)$ is motivated by the
isomorphism $\mathfrak{cl}(1,3)\simeq \mathfrak{cl}\left( 4\right) $,
whereas $\mathfrak{cl}(3,1)\simeq $ $\mathfrak{cl}(2,2)$. In the GA
formalism, the metric structure of the space whose geometric algebra is
built, reflects the properties of the unit pseudoscalar of the algebra.
Indeed, the existence of a pseudoscalar is equivalent to the existence of a
metric. In $\mathfrak{cl}(1,3)$ the highest-grade element, the unit
pseudoscalar, is defined as,%
\begin{equation}
i\overset{\text{def}}{=}\gamma _{0}\gamma _{1}\gamma _{2}\gamma _{3}\text{.}
\end{equation}%
It represents an oriented unit four-dimensional volume element. The
corresponding volume element is said to be right-handed because $i$ can be
generated from a right-handed vector basis by the oriented product $\gamma
_{0}\gamma _{1}\gamma _{2}\gamma _{3}$. The volume element $i$ has magnitude 
$\left\vert i\right\vert =\left\langle i^{\dagger }i\right\rangle _{0}^{%
\frac{1}{2}}=1$, where $\left\langle x\right\rangle _{0}$ denotes the
0-grade component of the multi-vector $x$. The dagger $\dagger $ is the 
\textit{reverse} or the \textit{Hermitian adjoint}.\textit{\ }For example,
given a multivector $x=\gamma _{1}\gamma _{2}$, $x^{\dagger }$ is obtained
by reversing the order of vectors in the product. That is, $x^{\dagger }=$%
\textit{\ }$\gamma _{2}\gamma _{1}=-\gamma _{1}\gamma _{2}$ . It is commonly
said that $i$ defines an orientation for spacetime. The pseudoscalar
satisfies $i^{2}=\pm 1$ with the sign depending on the dimension and the
signature of the space whose GA is considered. For instance, in spaces of
positive definite metric, the pseudoscalar has magnitude $\left\vert
i\right\vert =1$ while the value of $i^{2}$ depends only on the dimension of
space as $i^{2}=\left( -1\right) ^{n\left( n-1\right) /2}$. This implies
that in the positive-definite space\ of $SO(4)$ and in the zero signature
space-time of $SO(2,2)$, the unit pseudoscalar squares to $+1$. For the
space-time of the Lorentz group, the pseudoscalar satisfies $i^{2}=-1$.

Since we are dealing with a space of even dimension ($n=4$), $i$
anticommutes with odd-grade multi-vectors and \ commutes with even-grade
elements of the algebra,%
\begin{equation}
i\mathcal{P}=\pm \mathcal{P}i
\end{equation}%
where the multi-vector $\mathcal{P}$ is even for ($+$) and odd for ($-$).

An important spacetime vector that will be used in our formulation is the
spacetime vector derivative $\nabla $, defined by%
\begin{equation}
\nabla =\gamma ^{\mu }\partial _{\mu }\equiv \gamma ^{0}c^{-1}\partial
_{t}+\gamma ^{i}\partial _{i}\text{.}
\end{equation}%
By post-multiplying with $\gamma ^{0}$, we see that%
\begin{equation}
\nabla \gamma _{0}=c^{-1}\partial _{t}+\gamma ^{i}\gamma _{0}\partial
_{i}=c^{-1}\partial _{t}-\vec{\nabla}\text{,}
\end{equation}%
where $\overrightarrow{\nabla }$ is the usual vector derivative defined in
vector algebra. Similarly, multiplying the spacetime vector derivative by $%
\gamma ^{0}$, we get%
\begin{equation}
\gamma _{0}\nabla =c^{-1}\partial _{t}+\vec{\nabla}\text{.}
\end{equation}%
Finally, we notice that the spacetime vector derivative satisfies the
following relation:%
\begin{equation}
\square =\left( \gamma _{0}\nabla \right) \left( \nabla \gamma _{0}\right)
=c^{-2}\partial _{t}^{2}-\vec{\nabla}^{2}\text{,}
\end{equation}%
which is the d'Alembert operator used in the description of lightlike
traveling waves.

\section{EM interaction with electric charges and a magnetic monopole via
Proca fields : general case}

\subsection{Tensor Algebra Formalism}

In this section we consider a system of electric charges interacting with
magnetic monopoles where the interaction is mediated by massive photons or
Proca fields. In cgs units, the Lagrangian density describing such a
Maxwell-Proca (MP) system is given by%
\begin{equation}
\mathcal{L}_{MP}\left( A\right) =-\frac{1}{16\pi }\mathcal{F}_{\mu \nu }%
\mathcal{F}^{\mu \nu }+\frac{m_{\gamma }^{2}}{8\pi }A_{\mu }A^{\mu }-\frac{1%
}{c}J_{\mu }A^{\mu }\text{,}
\end{equation}%
where $m_{\gamma }=\frac{\omega }{c}$ is the inverse of the Compton length
associated with the photon mass of the field $A_{\mu }$, the current $J_{\mu
}$ is the electron 4-current, and the 4-vector potential $A_{\mu }$ is
associated with the magnetic charge of the monopole. The current decomposes
as $J_{\mu }\equiv \left( \rho \text{, }-\vec{j}\right) $. The field
strength is defined as $\mathcal{F}_{\mu \nu }=F_{\mu \nu }+G_{\mu \nu }$
where the electromagnetic field strength has the usual form $F_{\mu \nu
}=\partial _{\mu }A_{\nu }-\partial _{\nu }A_{\mu }$ and the monopole
contribution $G_{\mu \nu }$ is given by%
\begin{equation}
G^{\mu \nu }\left( x\right) =\frac{4\pi e_{m}}{c}\int d\tau dx^{\mu }\delta
^{(4)}\left( x-x_{monopole}\right) u^{\nu }\left( x\right) \text{,}
\end{equation}%
where $\tau $ is a timelike parameter and $u^{\nu }\left( x\right) =\frac{%
dx^{\nu }}{d\tau }$ denotes the velocity of the monopole. From the
Euler-Lagrange equations%
\begin{equation}
\frac{\partial \mathcal{L}_{MP}}{\partial A_{\mu }}-\partial _{\nu }\left( 
\frac{\partial \mathcal{L}_{MP}}{\partial \left( \partial _{\nu }A_{\mu
}\right) }\right) =0\text{,}
\end{equation}%
we obtain the field equation%
\begin{equation}
\partial _{\mu }F^{\mu \nu }+m_{\gamma }^{2}A^{\nu }=\frac{4\pi }{c}J^{\nu }%
\text{.}
\end{equation}%
By use of the Lorenz gauge condition%
\begin{equation}
\partial _{\mu }A^{\mu }=0\text{,}
\end{equation}%
(18) can be expressed explicitly in terms of the vector potential,%
\begin{equation}
\left( \square +m_{\gamma }^{2}\right) A_{\mu }=\frac{4\pi }{c}J_{\mu }\text{%
.}
\end{equation}%
Computing the Bianchi identity $\partial _{\mu }$ $^{\ast }F^{\mu \nu }$
where the Hodge dual is defined by $^{\ast }F^{\mu \nu }=\frac{1}{2}\epsilon
^{\mu \nu \alpha \beta }F_{\alpha \beta }$, we obtain%
\begin{equation}
\partial _{\mu }\text{ }^{\ast }F^{\mu \nu }=-\partial _{\mu }\text{ }G^{\mu
\nu }=-\frac{4\pi }{c}J_{(m)}^{\nu }\text{,}
\end{equation}%
where $J_{(m)}^{\nu }\equiv \left( \rho _{m}\text{, }\vec{j}_{(m)}\right) $, 
$\rho _{m}=e_{m}\delta ^{3}\left( \vec{x}-\vec{x}_{monopole}\right) $ and $%
\vec{j}_{(m)}=0$ since we are in the monopole rest frame.

By decomposing the field equation (18) into its boost and spatial components
and using (21), we arrive at the generalized Maxwell equations in presence
of massive photons and magnetic monopoles in the vector algebra formalism:%
\begin{equation}
\vec{\nabla}\cdot \vec{E}\mathbf{=}4\pi \rho _{e}-m_{\gamma }^{2}A_{0}\text{,%
}
\end{equation}%
\begin{equation}
\vec{\nabla}\times \vec{E}\mathbf{=}-c^{-1}\partial _{t}\vec{B}\text{,}
\end{equation}%
\begin{equation}
\vec{\nabla}\cdot \vec{B}\mathbf{=}4\pi \rho _{m}\text{,}
\end{equation}%
\begin{equation}
\vec{\nabla}\times \vec{B}\mathbf{=}4\pi c^{-1}\vec{j}_{e}+c^{-1}\partial
_{t}\vec{E}-m_{\gamma }^{2}\vec{A}\text{.}
\end{equation}%
Note that although the quantity $-4\pi c^{-1}\vec{j}_{m}$ is missing in (23)
(as it should since we are working in the rest frame of the monopole, $\vec{j%
}_{m}=0$) we will include this term in the following considerations for
purposes of generality.

\subsection{Spacetime Algebra Formalism}

We now recast the generalized vectorial equations (22)-(25) (after including 
$-4\pi c^{-1}\vec{j}_{m}$ in (23)) into the language of GA. We begin by
utilizing the wedge product (outer product) used in defining the geometric
product (1). By means of this operation, (23) and (25) become 
\begin{equation}
\vec{\nabla}\wedge \vec{E}\mathbf{=}-c^{-1}\partial _{t}(i\vec{B}\mathbf{)}%
-4\pi c^{-1}i\vec{j}_{m}\text{,}
\end{equation}%
\begin{equation}
\vec{\nabla}\wedge \vec{B}\mathbf{=}i\mathbf{(}4\pi c^{-1}\vec{j}%
_{e}+c^{-1}\partial _{t}\vec{E}-m_{\gamma }^{2}\vec{A}\mathbf{)}\text{.}
\end{equation}%
Adding the 0-grade (22) and the 2-grade (26) multi-vectorial equations in
which the electric field appears, and computing the equivalent quantity for
the magnetic field, we obtain%
\begin{equation}
\vec{\nabla}\vec{E}\mathbf{=}4\pi \rho _{e}-m_{\gamma
}^{2}A_{0}-c^{-1}\partial _{t}(i\vec{B})-4\pi c^{-1}i\vec{j}_{m}\text{,}
\end{equation}%
\begin{equation}
\vec{\nabla}(i\vec{B})\mathbf{=}-4\pi c^{-1}\vec{j}_{e}-c^{-1}\partial _{t}%
\vec{E}+m_{\gamma }^{2}\vec{A}+4\pi \rho _{m}i\text{,}
\end{equation}%
where in the last equation we used the fact that the pseudoscalar commutes
with the three dimensional vector derivative. Manipulating the two equations
in (28) and (29), we obtain%
\begin{equation}
(\vec{\nabla}+c^{-1}\partial _{t})F=4\pi c^{-1}(c\rho _{e}-\vec{j}_{e})+4\pi
c^{-1}i(c\rho _{m}-\vec{j}_{m})-m_{\gamma }^{2}A_{0}+m_{\gamma }^{2}\vec{A}%
\text{,}
\end{equation}%
where $F\overset{\text{def}}{=}\vec{E}+i\vec{B}=E^{i}\gamma _{i}\gamma
_{0}-B^{1}\gamma _{2}\gamma _{3}-B^{2}\gamma _{3}\gamma _{1}-$ $B^{3}\gamma
_{1}\gamma _{2}$ is the Faraday spacetime bi-vector. Furthermore, defining
the spacetime electric and magnetic current $j_{e}$ and\ $j_{m}$ as%
\begin{equation}
j_{e}\overset{\text{def}}{=}\left\{ 
\begin{array}{c}
j_{e}\cdot \gamma _{0}=c\rho _{e} \\ 
j_{e}\wedge \gamma _{0}=\vec{j}_{e}%
\end{array}%
\right\} \text{, }j_{m}\overset{\text{def}}{=}\left\{ 
\begin{array}{c}
j_{m}\cdot \gamma _{0}=c\rho _{m} \\ 
j_{m}\wedge \gamma _{0}=\vec{j}_{m}%
\end{array}%
\right\}
\end{equation}%
\ and by using equation (30) and the relations $\gamma _{0}i=-i\gamma _{0}$, 
$i^{2}=-1$ and $\gamma _{0}^{2}=1$ we arrive at%
\begin{equation}
\nabla F=4\pi c^{-1}(j_{e}-ij_{m})-m_{\gamma }^{2}(\gamma _{0}A_{0}-\gamma
_{0}\vec{A}\mathbf{)}\text{.}
\end{equation}%
Upon introduction of the spacetime vector potential $A$%
\begin{equation}
A\overset{\text{def}}{=}\left\{ 
\begin{array}{c}
A\cdot \gamma _{0}=A_{0} \\ 
A\wedge \gamma _{0}=\vec{A}%
\end{array}%
\right\} \text{,}
\end{equation}%
(32) takes the final form%
\begin{equation}
\nabla F=4\pi c^{-1}(j_{e}-ij_{m})-m_{\gamma }^{2}A\text{.}
\end{equation}%
This equation represents the GA formulation of the fundamental equations of
massive classical electrodynamics in presence of magnetic monopoles. This
equation can be decomposed into its vectorial and tri-vectorial components,
namely%
\begin{equation}
\nabla \cdot F=4\pi c^{-1}j_{e}-m_{\gamma }^{2}A\text{,}
\end{equation}%
\begin{equation}
\nabla \wedge F=-4\pi c^{-1}ij_{m}\text{.}
\end{equation}%
From equation (34), we determine%
\begin{eqnarray}
\nabla ^{2}F &=&4\pi c^{-1}\nabla \cdot j_{e}+4\pi c^{-1}\nabla \wedge
j_{e}+4\pi c^{-1}i\nabla \cdot j_{m} \\
&&+4\pi c^{-1}i\nabla \wedge j_{m}-m_{\gamma }^{2}\nabla \cdot A-m_{\gamma
}^{2}\nabla \wedge A\text{.}  \notag
\end{eqnarray}%
By splitting this equation in its different multi-vectorial parts, we have%
\begin{equation}
\nabla ^{2}F=4\pi c^{-1}(\nabla \wedge j_{e}+i\nabla \wedge j_{m})-m_{\gamma
}^{2}\nabla \wedge A\text{,}
\end{equation}%
\begin{equation}
4\pi c^{-1}\nabla \cdot j_{e}-m_{\gamma }^{2}\nabla \cdot A=0\text{,}
\end{equation}%
\begin{equation}
4\pi c^{-1}i\nabla \cdot j_{m}=0\text{.}
\end{equation}%
Equation (39) implies%
\begin{equation}
\nabla \cdot j_{e}=\frac{c}{4\pi }m_{\gamma }^{2}\nabla \cdot A\text{,}
\end{equation}%
while from (40) we get%
\begin{equation}
i\nabla \cdot j_{m}=0\text{.}
\end{equation}%
In order to maintain charge conservation $\nabla \cdot j_{e}=0$, since $%
m_{\gamma }\neq 0$, the Lorenz gauge condition $\nabla \cdot A=0$ must be
satisfied.

\section{EM interaction with electric and magnetic charges via Proca fields:
special cases}

We consider two limiting cases of the GA algebra formulation of massive
classical electrodynamics in presence of magnetic monopoles: massive
electrodynamics in absence of monopoles and alternatively, standard Maxwell
electrodynamics with monopoles.

\subsection{Electrodynamics with massive photons}

Let's consider the limiting case of equation (34) when no monopoles are
present, that is%
\begin{equation}
\nabla F=4\pi c^{-1}j_{e}-m_{\gamma }^{2}A\text{.}
\end{equation}%
Considering the tri-vector and vector parts of (43), we get%
\begin{equation}
\nabla \wedge F=0\text{,}
\end{equation}%
\begin{equation}
\nabla \cdot F=4\pi c^{-1}j_{e}-m_{\gamma }^{2}A\text{.}
\end{equation}%
In the tensor algebra formalism, equation $\left( 44\right) $ becomes the
tensorial relation $\partial _{\lambda }F_{\mu \nu }+\partial _{\nu
}F_{\lambda \mu }+\partial _{\mu }F_{\nu \lambda }=0$. Moreover, equation $%
(44)$ implies that the Faraday bi-vector can be written as an outer
derivative of a certain dynamical vector variable $A$ which couples to the
external spacetime electric current $j_{e}$,%
\begin{equation}
F=\frac{1}{2}F^{\mu \nu }\gamma _{\mu }\wedge \gamma _{\nu }=\nabla \wedge A
\end{equation}%
where $F^{\mu \nu }=\gamma ^{\mu }\wedge \gamma ^{\nu }\cdot F$ are the
components of $F$ in the $\left\{ \gamma ^{\mu }\right\} $ frame.
Substituting this expression of $F$ in equation $(45)$, we get the following
equation for $A$,%
\begin{equation}
\nabla ^{2}A-\nabla (\nabla \cdot A)=4\pi c^{-1}j_{e}-m_{\gamma }^{2}A\text{.%
}
\end{equation}%
It is clear that this equation is not invariant under the standard local
gauge transformation $A\rightarrow A+\nabla \chi (x)$, which implies a loss
of local gauge invariance in massive classical electrodynamics. This point
will be discussed further in section 5.1.

\subsection{Electrodynamics with magnetic charges}

Let us consider now the limiting case of equation (34), when the
electromagnetic interaction is mediated by massless photons and magnetic
monopoles are allowed to occur in the theory. This situation is reflected in
the relation%
\begin{equation}
\nabla F=4\pi c^{-1}(j_{e}-ij_{m})\text{.}
\end{equation}%
Equation (48) is equivalent to the following pair of equations%
\begin{equation}
\nabla \cdot F=4\pi c^{-1}j_{e}\text{,}
\end{equation}%
\begin{equation}
\nabla \wedge F=-4\pi c^{-1}ij_{m}\text{.}
\end{equation}%
This last equation implies the vectorial fields $\overrightarrow{E}$\textbf{%
\ }and $\overrightarrow{B}$\textbf{\ }can no longer be described in terms of
the dynamical quantity $A$. Furthermore, equation (48) leads to%
\begin{equation}
\nabla ^{2}F=4\pi c^{-1}\nabla j_{e}-4\pi c^{-1}\nabla (ij_{m})=4\pi
c^{-1}\nabla j_{e}+4\pi c^{-1}i\nabla j_{m}
\end{equation}%
that is,%
\begin{equation}
\nabla ^{2}F=4\pi c^{-1}\nabla \cdot j_{e}+4\pi c^{-1}\nabla \wedge
j_{e}+4\pi c^{-1}i\nabla \cdot j_{m}+4\pi c^{-1}i\nabla \wedge j_{m}\text{.}
\end{equation}%
By splitting the different multi-vectorial parts of (52) we find%
\begin{equation}
\nabla ^{2}F=4\pi c^{-1}\nabla \wedge j_{e}+4\pi c^{-1}i\nabla \wedge j_{m}%
\text{,}
\end{equation}%
\begin{equation}
\nabla \cdot j_{e}=0\text{,}
\end{equation}%
\begin{equation}
i\nabla \cdot j_{m}=0\text{.}
\end{equation}%
It is clear from equations (54) and (55) that charge conservation is
satisfied.

\section{Invariances of the EM Lagrangian and Hamiltonian densities}

We now consider the spacetime \textit{duality rotations} (DR) and local
gauge invariances of the Lagrangian $\mathcal{L}$ and Hamiltonian $\mathcal{H%
}$ densities. It is demonstrated that $\mathcal{L}$ (as well as $\mathcal{H}$%
) looses its local gauge invariance due to the presence of the Proca mass
term. Furthermore, $\mathcal{L}$ is shown to be non-invariant under DR. The
latter result is well known $[7]$. In the conventional approach, this
reflects the fact that the Lagrangian is an indefinite quantity, that is,
there is a relative sign difference between the kinetic and potential energy
terms. Since $\left( \vec{E},\vec{B}\right) \overset{\text{DR}}{\rightarrow }%
\left( -\vec{B},\vec{E}\right) $ for a rotation of $\alpha =\frac{\pi }{2}$,
the Faraday bi-vector transforms as $F\overset{\text{DR}}{\rightarrow }-iF$
and one would expect a change in the relative sign of $\mathcal{L}$. Hence,
in the free-field electromagnetic case, $\mathcal{L}$ would transform as $%
\pm \mathcal{L}\overset{\text{DR}}{\rightarrow }\mathcal{L}^{\prime }=\mp 
\mathcal{L}$. In contrast, since $\mathcal{H}$ is positive definite, $F%
\overset{\text{DR}}{\rightarrow }-iF$ so that $\mathcal{H}^{\prime }$ is
also positive definite, where $\mathcal{H}\overset{\text{DR}}{\rightarrow }%
\mathcal{H}^{\prime }=\mathcal{H}$.

This being the case, it is interesting to consider the following. The
transformation properties of $\mathcal{L}$ and $\mathcal{H}$ under DR is
consistent with the fact that in the GA of spacetime, the unit 4-volume
element $i$ squares to $-1$. It thus appears plausible to view the DR
invariances of $\mathcal{L}$ and $\mathcal{H}$ as a manifestation of
spacetime topology. This point will be further discussed in 5.2. Finally, in
accordance with the above arguments $\mathcal{H}$ is shown to be invariant
under duality rotations.

\subsection{Gauge Invariance}

Consider the Maxwell-Proca (MP) Lagrangian density (15) expressed in the GA
formalism%
\begin{equation}
\mathcal{L}_{MP}^{\left( GA\right) }\left( A\right) =\alpha _{1}\left\langle
F\cdot F\right\rangle _{0}+\alpha _{2}\left\langle A\cdot J\right\rangle
_{0}+\alpha _{3}m_{\gamma }^{2}\left\langle A\cdot A\right\rangle _{0}\text{,%
}
\end{equation}%
where $\alpha _{i}$ are real constants whose values are not specified in our
discussion. The Faraday spacetime bi-vector is given by $F=\nabla \wedge A$.
The second term of $\mathcal{L}_{MP}^{\left( GA\right) }\left( A\right) $
describes the coupling between $A$ and the external current $J$. The last
term represents the mass term. Let us discuss the invariance of $\mathcal{L}%
_{MP}^{\left( GA\right) }\left( A\right) $ under Local Gauge Transformation
(LGT) defined by%
\begin{equation}
A\overset{\text{LGT}}{\longrightarrow }A^{\prime }=A+\nabla \chi (x)\text{,}
\end{equation}%
where $\chi (x)$ is the scalar gauge field. For the first term we find%
\begin{equation}
F\overset{\text{LGT}}{\longrightarrow }F^{\prime }=\nabla \wedge A^{\prime
}=\nabla \wedge \left( A+\nabla \chi (x)\right) =\nabla \wedge A+\nabla
\wedge \left( \nabla \chi (x)\right) =F\text{.}
\end{equation}%
The first term of $\mathcal{L}_{MP}^{\left( GA\right) }\left( A\right) $ is
gauge invariant since%
\begin{equation}
\left\langle F\cdot F\right\rangle _{0}\overset{\text{LGT}}{\longrightarrow }%
\left\langle F^{\prime }\cdot F^{\prime }\right\rangle _{0}=\left\langle
F\cdot F\right\rangle _{0}\text{,}
\end{equation}%
while for the second term,%
\begin{equation}
A\cdot J\rightarrow A^{\prime }\cdot J=A\cdot J-\chi \nabla \cdot J+\nabla
\cdot \left( \chi J\right) \text{.}
\end{equation}%
Since $\nabla \cdot \left( \chi J\right) $ is a total divergence it can be
ignored because its integral over a finite volume results in a boundary term
which can be set to zero. Finally, the gauge invariance of this term is
ensured by requiring that the external current $J$ is conserved, $\nabla
\cdot J=0$. We now turn to the last term in $\mathcal{L}_{MP}^{\left(
GA\right) }\left( A\right) $. Under local gauge transformation,%
\begin{equation}
A\cdot A\overset{\text{LGT}}{\longrightarrow }A^{\prime }\cdot A^{\prime
}=A^{2}+\left( \nabla \chi \right) ^{2}+2\nabla \cdot \left( \chi A\right)
-2\chi \nabla \cdot A\text{.}
\end{equation}%
Using the Lorentz condition $\nabla \cdot A=0$ and ignoring $\nabla \cdot
\left( \chi A\right) $ for reasons described above, we obtain,%
\begin{equation}
A\cdot A\overset{\text{LGT}}{\longrightarrow }A^{\prime }\cdot A^{\prime
}=A^{2}+\left( \nabla \chi \right) ^{2}\text{.}
\end{equation}%
Clearly, this term is not gauge invariant due to the occurrence of the
non-vanishing square gradient $\left( \nabla \chi \right) ^{2}$. Local gauge
symmetry is therefore lost in electromagnetic interactions mediated by
massive photons.

\subsection{EM Duality and 4D spacetime signature}

Two fundamental properties characterize the physical space: its signature
and its dimensionality $D=4$ $[8]$. An explanation of why spacetime has $3+1$
signature rather than $4+0$ or $2+2$ metric is discussed in $[9]$.

It is shown in $[10]$ that there is a relation between four signs in
electrodynamics and the signature: i) the "$-$" sign in the Ampere-Maxwell
law, $\vec{\nabla}\times \vec{B}\mathbf{-}c^{-1}\partial _{t}\vec{E}=0$ ;
ii) the "$+$" sign in the Faraday law (Lenz rule), $\vec{\nabla}\times \vec{E%
}\mathbf{+}c^{-1}\partial _{t}\vec{B}=0$ ; iii) the "$+$" sign in the
electromagnetic density, $u_{EM}=\frac{1}{8\pi }\left( \vec{E}^{2}+\vec{B}%
^{2}\right) $; iv) the $(+$,$-$,$-$,$-)$ signature of the Lorentz metric. In 
$[10]$ it is also shown that given a manifold endowed with a certain
signature, the electric and magnetic energy densities are positive in the
Minkowskian case, while they have opposite sign in the Euclidean
electrodynamics $[11,12]$. These results will be briefly discussed in the GA
formalism and, in addition, it will be shown that the EM duality invariance
of the Lagrangian and Hamiltonian densities in the Minkowskian and Euclidean
cases are related to the signature of the metric of the manifold over which
the dynamics is constructed.

\subsubsection{The Minkowskian case}

For the sake of simplicity, assume $c=1$. Consider the Lenz law with "$+$"
sign,%
\begin{equation}
\vec{\nabla}\times \vec{E}+\frac{\partial \vec{B}}{\partial t}=0
\end{equation}%
We will show that the "$+$" sign will lead to the correct relativistic wave
equation. Considering the curl of (63) and introducing the outer product "$%
\wedge $", we obtain%
\begin{equation}
\vec{\nabla}\times \left( \vec{\nabla}\times \vec{E}\right) =i_{\mathbf{M}%
}^{2}\nabla \cdot \left( \vec{\nabla}\wedge \vec{E}\right) =-\frac{\partial 
}{\partial t}\left( \vec{\nabla}\times \vec{B}\right)
\end{equation}%
where the subscript $\mathbf{M}$ represents Minkowski spacetime. In absence
of sources, $\vec{\nabla}\cdot \vec{E}=0$, and recalling that $\vec{\nabla}%
\times \vec{B}-\frac{\partial \vec{E}}{\partial t}=0$, we obtain%
\begin{equation}
\square _{\mathbf{M}}\vec{E}\equiv \left( i_{\mathbf{M}}^{2}\vec{\nabla}^{2}+%
\frac{\partial ^{2}}{\partial t^{2}}\right) \vec{E}=0
\end{equation}%
where $i_{\mathbf{M}}=\gamma _{0}\gamma _{1}\gamma _{2}\gamma _{3}$ is the
Minkowski spacetime unit pseudoscalar that satisfies $i_{\mathbf{M}}^{2}=-1$%
. The GA formalism emphasizes the fact that the shape of the second order
differential operator $\square _{\mathbf{M}}$ reflects the signature of the
metric. Moreover, in the GA formalism the Lagrangian density for
electromagnetism in the absence of sources is proportional to%
\begin{equation}
^{\mathbf{M}}\mathcal{L}_{\gamma }^{(GA)}\propto \left\langle
F^{2}\right\rangle _{0}=\vec{E}^{2}-\vec{B}^{2}\text{,}
\end{equation}%
where,%
\begin{equation}
F^{2}=\left\langle F^{2}\right\rangle _{0}+\left\langle F^{2}\right\rangle
_{4}=\vec{E}^{2}-\vec{B}^{2}+2i_{\mathbf{M}}\vec{E}\cdot \vec{B}\text{.}
\end{equation}%
Notice that $\left\langle F^{2}\right\rangle _{0}$ and $\left\langle
F^{2}\right\rangle _{4}$ are Lorentz invariant terms. Under Minkowski
Duality Rotations (MDR) with an arbitrary angle $\alpha $, the \textit{%
Faraday spacetime bi-vector\ }$F=\vec{E}+i_{\mathbf{M}}\vec{B}$ transforms
as 
\begin{equation}
F\longrightarrow F^{\prime }=Fe^{-i_{\mathbf{M}}\alpha }\text{.}
\end{equation}%
For the special case $\alpha =\frac{\pi }{2}$, $(\vec{E},\vec{B})\overset{%
\text{MDR}}{\rightarrow }(\vec{B},-\vec{E})$, $F$ transforms as%
\begin{equation}
F\overset{\text{MDR}}{\longrightarrow }F^{\prime }=-i_{\mathbf{M}}F\text{.}
\end{equation}%
Thus%
\begin{equation}
\left\langle F^{2}\right\rangle _{0}\overset{\text{MDR}}{\longrightarrow }%
\left\langle F^{\prime 2}\right\rangle _{0}=i_{\mathbf{M}}^{2}\left\langle
F^{2}\right\rangle _{0}=-\left\langle F^{2}\right\rangle _{0}\text{,}
\end{equation}%
where we used the fact that the spacetime unit volume $i_{\mathbf{M}}$ (the
unit pseudoscalar) squares to $-1$. Finally, from (62) and (66) we conclude
that under MDR $^{\mathbf{M}}\mathcal{L}_{\gamma }^{(GA)}$ changes its sign
as,%
\begin{equation}
\mathcal{L}_{\gamma }^{(GA)}\overset{\text{MDR}}{\longrightarrow }-\mathcal{L%
}_{\gamma }^{(GA)}\text{.}
\end{equation}%
The significant point here is that \textit{electromagnetic duality} is a 
\textit{symmetry} that is not exhibited as an \textit{invariance} of the 
\textit{Lagrangian density} and moreover, in the GA formalism this reflects
the signature of spacetime; or more specifically the fact that the the unit
pseudoscalar squares to $-1$. It can be easily shown that the free-field
electromagnetic energy-momentum tensor $T\left( a\right) =-\frac{1}{2}FaF$,
where $a=a^{\mu }\gamma _{\mu }$, is either gauge-invariant or invariant
under MDR. Furthermore, notice that $\left\langle F^{2}\right\rangle _{4}$
is not invariant under MDR.

On the other hand, the Hamiltonian density is proportional to%
\begin{equation}
^{\mathbf{M}}\mathcal{H}_{\gamma }^{(GA)}\varpropto \left\langle FF^{\dagger
}\right\rangle _{0}=\vec{E}^{2}+\vec{B}^{2}\text{,}
\end{equation}%
where%
\begin{equation}
FF^{\dagger }=\left\langle FF^{\dagger }\right\rangle _{0}+\left\langle
FF^{\dagger }\right\rangle _{2}=\vec{E}^{2}+\vec{B}^{2}-2i_{\mathbf{M}}\vec{E%
}\wedge \vec{B}
\end{equation}%
and the dagger $\dagger $ is the \textit{reverse} or the \textit{Hermitian
adjoint}.\textit{\ }For example, given a multivector $A=\gamma _{1}\gamma
_{2}$, $A^{\dagger }$ is obtained by reversing the order of vectors in the
product. That is, $A^{\dagger }=$\textit{\ }$\gamma _{2}\gamma _{1}=-\gamma
_{1}\gamma _{2}$. Notice that $\left\langle FF^{\dagger }\right\rangle _{0}$
is not Lorentz invariant. Under MDR\textit{\ }with $\alpha =\frac{\pi }{2}$,
we have%
\begin{equation}
\left\langle FF^{\dagger }\right\rangle _{0}\overset{\text{MDR}}{%
\longrightarrow }\left\langle F^{\prime }F^{\prime \dagger }\right\rangle
_{0}=-i_{\mathbf{M}}^{2}\left\langle FF^{\dagger }\right\rangle
_{0}=\left\langle FF^{\dagger }\right\rangle _{0}\text{.}
\end{equation}%
Therefore, from (68) and (70) the Hamiltonian density $^{\mathbf{M}}\mathcal{%
H}_{\gamma }^{(GA)}$ is invariant under electromagnetic duality rotation:%
\begin{equation}
^{\mathbf{M}}\mathcal{H}_{\gamma }^{(GA)}\overset{\text{MDR}}{%
\longrightarrow }\text{ }^{\mathbf{M}}\mathcal{H}_{\gamma }^{(GA)}\text{.}
\end{equation}%
Notice that $\left\langle FF^{\dagger }\right\rangle _{2}\equiv \frac{8\pi }{%
c}\vec{S}$ , where $\vec{S}$ is the Poynting vector, is invariant under MDR
also. In view of the meaning of the oriented 4-volume element $i$, and
considering the transformation properties of the Hamiltonian density under
MDR, we are lead to conclude that, \textit{the EM-duality invariance of the
free-field electromagnetic Hamiltonian density reflects the signature of the
spacetime whose GA is built}. The duality invariance of $^{\mathbf{M}}%
\mathcal{H}_{\gamma }^{(GA)}$ is fundamental in order to preserve the
positive definiteness of energy density.

Phase and spacetime duality rotations are transformations that occur on
different spaces: the former correspond to internal gauge rotations in the
canonical phase space while the latter are external (spacetime) Lorentz
transformations (LT). We note however that this MDR ($\alpha =\pi /2$) can
be obtained as special cases of LT. By requiring the quantities $\vec{E}%
\cdot \vec{B}=\vec{E}^{\prime }\cdot \vec{B}^{\prime }$ and $\vec{E}^{2}-%
\vec{B}^{2}=\vec{E}^{\prime 2}-\vec{B}^{\prime 2}$ be invariant, it follows
that $\left( \vec{E}\text{,}\vec{B}\right) \overset{\text{LT}}{\rightarrow }%
\left( \vec{E}^{\prime }\text{,}\vec{B}^{\prime }\right) $ such that $%
E_{i}B_{j}\delta _{ij}=0$ and $E_{i}E_{j}\delta _{ij}=B_{i}B_{j}\delta _{ij}$
mimics the desired DT. This suggests a link between spacetime and gauge
symmetries. Indeed, in the canonical formulation, a duality phase rotation
emerges in $^{\mathbf{M}}\mathcal{L}\rightarrow $ $^{\mathbf{M}}\mathcal{L}%
^{\prime }=e^{-2i_{\mathbb{C}}\phi }$ $^{\mathbf{M}}\mathcal{L}$.
Considering the GA equivalent $^{\mathbf{M}}\mathcal{L}^{(GA)}\overset{%
\alpha =\pi /2}{\longrightarrow }$ $^{\mathbf{M}}\mathcal{L}^{\prime
(GA)}=e^{-2i_{\mathbf{M}}\alpha }$ $^{\mathbf{M}}\mathcal{L}^{(GA)}$, we
arrive at the \emph{correspondence }$e^{-2i_{\mathbb{C}}\phi }\sim i_{%
\mathbf{M}}^{2}$. This link between spacetime signature $i_{\mathbf{M}}^{2}$
and gauge phase factor $e^{-2i_{\mathbb{C}}\phi }$ at this stage is purely
mathematical. The geometrization program of physics is not sufficiently
advanced to provide an adequate interpretation of gauge symmetry. An
understanding of any potential link between these two types of invariances
requires a deeper geometrical analysis of local gauge invariance. For this
purpose, the GA formalism seems to be most adequately suited.

\subsubsection{The Euclidean case}

The Euclidean Maxwell equations in absence of sources in flat space are $%
[12] $,%
\begin{equation}
\vec{\nabla}\cdot \vec{E}=0\text{, }\vec{\nabla}\cdot \vec{B}=0\text{, }\vec{%
\nabla}\times \vec{E}-\frac{\partial \vec{B}}{\partial t}=0\text{, }\vec{%
\nabla}\times \vec{B}-\frac{\partial \vec{E}}{\partial t}=0\text{.}
\end{equation}%
The remarkable difference from the usual vacuum Maxwell equations is that
there is an "anti-Lenz" law, with "$+$" sign replaced by a "$-$" sign. A
main consequence of such a change is that in Euclidean 4-spaces there is no
propagation with finite speed, 
\begin{equation}
\square _{\mathbf{E}}\vec{E}\equiv \left( i_{\mathbf{E}}^{2}\vec{\nabla}^{2}+%
\frac{\partial ^{2}}{\partial t^{2}}\right) \vec{E}=0\text{,}
\end{equation}%
where the subscript $\mathbf{E}$ represents Euclidean 4-space, $i_{\mathbf{E}%
}=e_{0}e_{1}e_{2}e_{3}$ is the 4-space Euclidean unit pseudoscalar
satisfying $i_{\mathbf{E}}^{2}=1$. The multivectors $e_{j}$ satisfy%
\begin{equation}
e_{0}^{2}=1\text{, }e_{0}\cdot e_{i}=0\text{ and }e_{i}\cdot e_{j}=\delta
_{ij}\text{; }i\text{, }j=0..3\text{.}
\end{equation}%
The monochromatic wave solutions of (77) are either exponentially growing or
decaying as a function of distance along the direction of propagation. It is
clear, using the GA formalism that the unit pseudoscalar of the algebra
encodes information about the structure of the differential operator
describing wave propagation in the spaces whose GA is constructed. Finally,
notice that the Euclidean Maxwell equations in (76) are invariant under the
Euclidean Duality Rotation (EDR) $(\vec{E},\vec{B})\overset{\text{EDR}}{%
\rightarrow }(\vec{B},\vec{E})$. In terms of the Euclidean Faraday bi-vector 
$\tciFourier =\vec{E}+i_{\mathbf{E}}\vec{B}$, the Lagrangian density becomes 
\begin{equation}
^{\mathbf{E}}\mathcal{L}_{\gamma }^{(GA)}\propto \left\langle \tciFourier
^{2}\right\rangle _{0}=\vec{E}^{2}+\vec{B}^{2}\text{,}
\end{equation}%
where $\tciFourier ^{2}=\vec{E}^{2}+\vec{B}^{2}+2i_{\mathbf{E}}\vec{E}\cdot 
\vec{B}$. Also, the Hamiltonian density is,%
\begin{equation}
^{\mathbf{E}}\mathcal{H}_{\gamma }^{(GA)}\propto \left\langle \tciFourier
\tciFourier ^{\dagger }\right\rangle _{0}=\vec{E}^{2}-\vec{B}^{2}\text{,}
\end{equation}%
where $\tciFourier \tciFourier ^{\dagger }=\vec{E}^{2}-\vec{B}^{2}-2i_{%
\mathbf{E}}\vec{E}\wedge \vec{B}$. The GA formalism emphasizes the fact that
the definiteness of the Hamiltonian and Lagrangian densities are related to
the signature of the metric manifold and that only the positive-definite
quantities preserve the duality invariance.

\section{Conclusions \ }

Spacetime algebra is used to unify Maxwell's equations with non-vanishing
photon mass and Dirac monopoles. The theory is described by a
non-homogeneous multi-vectorial equation. The physical content of the theory
is not obscured by reference to specific choices of frames or set of
coordinates. Moreover, gauge and spacetime invariances of the theory have
been considered in the GA formalism. In addition to reproducing all standard
invariances, it is shown that the GA approach also supports the idea of an
EM origin for the Lorentzian signature of spacetime. It is shown that the EM
duality invariance of the Lagrangian and Hamiltonian densities in the
Minkowskian and Euclidean cases are related to the signature of the metric
of the manifold over which the dynamics is constructed. Moreover, GA
formalism makes clear that DR are symmetries of the definite quantities in
the Minkowskian and Euclidean manifolds, namely Hamiltonian and Lagrangian
densities respectively.

\section{Acknowledgements}

We are indebted to Dr. J. Kimball for concretely contributing to the
improvment of this manuscript. We also thank Dr. A. Inomata and Dr. A.
Caticha for useful discussions. Finally, we are grateful to P. Young for his
encouragement during this work.

\end{document}